\documentclass[twocolumn,showpacs,superscriptaddress,groupedaddress,floatfix]{revtex4-1}
\usepackage{color}
\usepackage[]{graphicx}
\usepackage{bm}        
\usepackage{amssymb}
\usepackage{amsmath}   
\usepackage{float}
\usepackage[english]{babel}

\begin{document}

\title{Cooling of Levitated Graphene Nanoplatelets in High Vacuum}
\author{Pavel Nagornykh}
\affiliation{Department of Physics, University of Maryland, College Park, Maryland 20742, USA}
\author{Joyce E. Coppock}
\affiliation{Department of Physics, University of Maryland, College Park, Maryland 20742, USA}
\author{B. E. Kane}
\affiliation{Joint Quantum Institute, University of Maryland, College Park, Maryland 20742, USA}
\affiliation{Laboratory for Physical Sciences,  College Park, Maryland 20740, USA}
\begin{abstract}
We demonstrate cooling of the center of mass motion of charged graphene nanoplatelets levitated in a quadrupole ion trap in high vacuum down to temperatures of 20 K. Parametric feedback based on optical measurements of particle motion was used to achieve the particle cooling at pressure $p < 10^{-6}$ Torr, and cooling along all three axes of motion was observed. Dependence of cooling on the electric fields was measured by varying DC voltages on a set of auxiliary electrodes used to spatially shift the trap minimum. Methods to calibrate mass and charge of the nanoplatelet by measuring its motion frequency dependence on discharge were also explored.
\end{abstract}

\maketitle

\begin{figure}
\includegraphics[width=\linewidth]{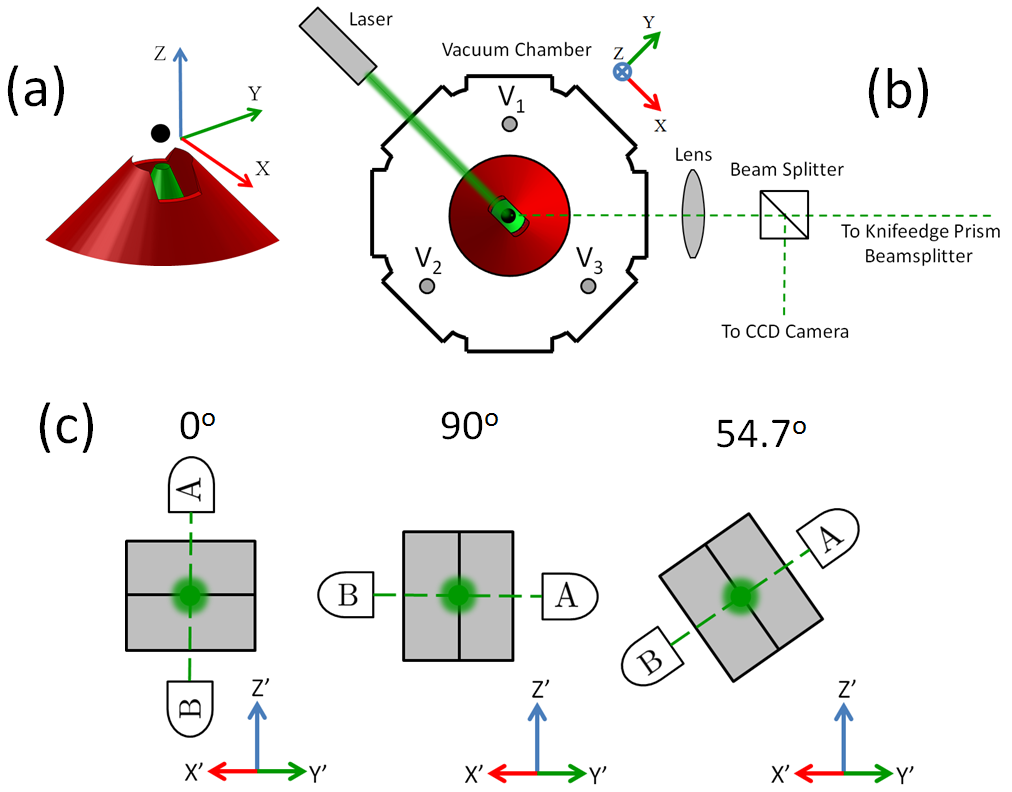}
\caption{\label{experiment} Diagram of experimental apparatus: (a) The particle trap consists of an outer electrode with a slot (red) and an inner electrode (green); (b) Schematic setup of the experiment as viewed from above the trap. Inside the high vacuum chamber auxiliary electrodes ($V_1$, $V_2$, $V_3$) are used to null stray electric field; (c) View of the prism mirror beam splitter as seen along the direction of the light scattered from the particle.~$X'$, $Y'$ and $Z'$ show the projections of the main axis of motion of the particle onto the lens' focal plane for three different orientations of the prism beam splitter. Photodectors are shown schematically as $A$ and $B$.}
\end{figure}

\begin{figure}
    \includegraphics[width=0.7\linewidth]{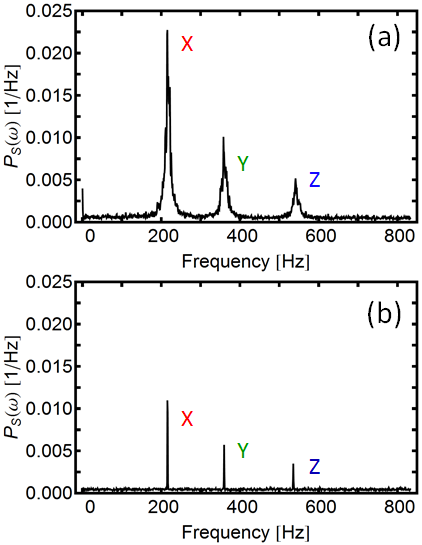}
    \caption{\label{PowerSpectrum} Power spectrum of signal $S$ at (a) $p = 4$ mTorr and (b) $p = 4\times 10^{-7}$ Torr.}
\end{figure}

\begin{figure}
\includegraphics[width=0.68\linewidth]{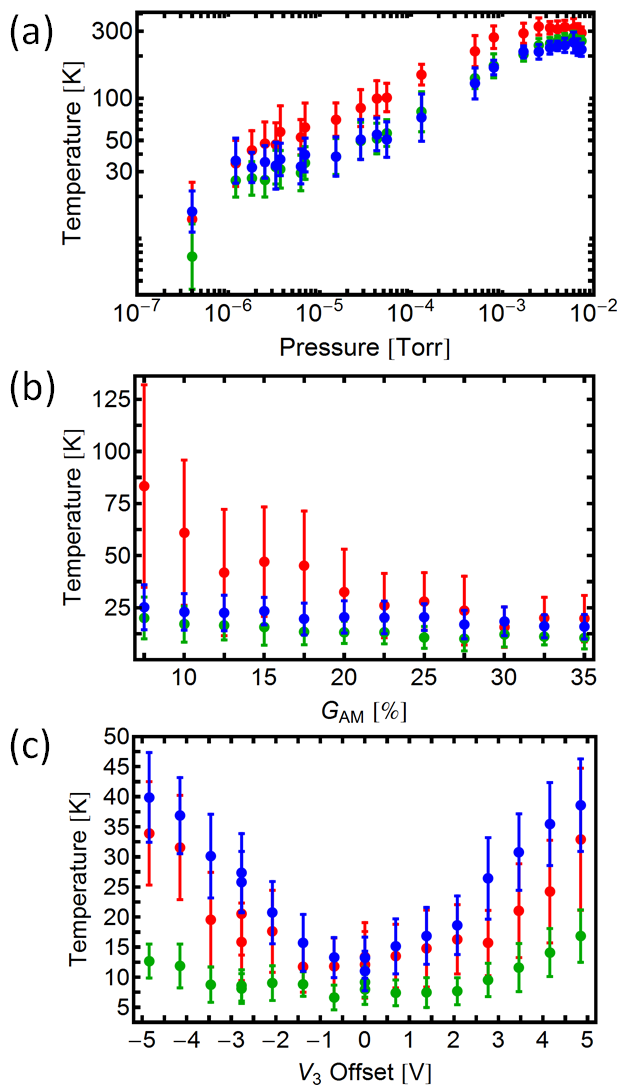}
\caption{\label{PressureAMVoltage} Dependence of $T$ of the center of mass motion along the trap axes $X$ (red), $Y$ (green) and $Z$ (blue) on: (a) $p$ ; (b) amplitude modulation $G_{\mathrm{AM}}$ [at $p = 4\times 10^{-7}$ Torr]; (c) voltage $V_3$ (values plotted relative to the nulling setting).}
\end{figure}

While trapping and cooling technologies have been primarily applied to atomic systems~\cite{Wineland1998,Leibfried2003,Kaufmann2012}, they have also found applications in the study of nanoscale systems of condensed matter \cite{Li2011,Guan2011,Millen2014,Marago2013,Kiesel27082013,Asenbaum2013,Kuhlicke2014}. One of many 2D materials that could benefit from these techniques is graphene, a material that garnered attention after its discovery in 2004~\cite{Novoselov2004}. A plethora of new phenomena related to unique behavior of electrons in graphene were predicted~\cite{Geim2009,CastroNeto2009}. Although substantial progress has been made in getting closer to graphene's intrinsic behavior~\cite{Dean2010,Bolotin2008,Meyer2007}, there are certain properties that are hard to measure in conventional experiments, and consequently a complementary approach of decoupling of graphene from the environment by levitating small graphene samples was proposed~\cite{Kane2010}. This experimental setup has the advantage of allowing the study of free graphene membrane motion at ultrahigh rotation speeds, which can be used to apply strain to the membrane~\cite{Kane2010,Arita2013}. Thermal properties of graphene near its melting point, predicted to be close to temperature $T = 4510$\,K~\cite{Los2015}, are also out of reach for standard measurements while easily achievable by laser heating~\cite{Millen2014}. Finally, the levitation approach can be used to conduct research on ultra high vacuum (UHV) crystal growth, graphene manipulation and deposition onto a substrate as well as flakes' functionalization~\cite{Cocchi2012,Georgakilas2012}.
\\
\indent
In previous experiments~\cite{Kane2010}, a quadrupole ion trap was used to levitate graphene nano-platelets separated by a distance of about 2 mm from the trap electrodes. During the experiment, the time dependence of the center of mass motion of the particle and its visible intensity were measured at different pressure~($p$) levels, and circularly polarized light was used to provide spin to the particle. While these measurements produced estimates of the mass, thickness, and charge, it was also found that the time particles stay trapped before they escape was limited to a few hours at $p < 10^{-7}$~Torr~\cite{Kane2010}. However, lower $p$ values are required to obtain a low contamination environment necessary for material deposition, thermal experiments, and high spinning frequency measurements on graphene flakes.~Consequently, a solution to the problem of low trap life times is required.
\\
\indent
A similar problem has been known to exist for optically trapped microscopic particles, and parametric feedback cooling was developed as a possible solution~\cite{Novotny2012}. Experiments conducted in Ref.~\cite{Novotny2012} involved silica nanospheres trapped in an optical trap and cooling to 50 mK was observed. In this letter, we discuss our implementation of the parametric feedback approach for graphene flakes in the quadrupole ion trap, which results in improvement of the trapping times of the flakes compared to the non-cooled case, as well as flakes' stabilization and cooling in high vacuum conditions.
\label{sec:setup}
\\
\indent
The experimental setup used for our measurements is similar to the design used in Ref.~\cite{Kane2010} with modifications related to transfer of the particles after trapping and the way their motion is detected. The system consists of two chambers separated by a gate valve, with ionized particles introduced into the loading chamber by use of the electrospray technique~\cite{Fenn1990}. After trapping in the loading chamber, particles are transferred to the high vacuum chamber (see Fig.~\ref{experiment}.b), where the cooldown experiment is conducted. The use of two chambers minimizes contamination of the high vacuum chamber by volatile substances that are created during electrospray operation.
\\
\indent
When trapped in the high vacuum chamber, charged particles are levitated in an electric quadrupole trap~\cite{Paul1990} that consists of two coaxial conical electrodes (Fig.~\ref{experiment}.a). The inner electrode of the trap is kept at zero voltage while an AC voltage $V_{AC}^0=300$ V at frequency $\Omega/2\pi$ (chosen to allow stable trapping of particles) is applied to the outer electrode. The AC voltage is created by a high voltage amplifier with a signal generator connected to its input. Near the trap minimum, the trap potential is approximated by~\cite{Paul1990}:
\begin{equation}
\label{Vac}
V(t) = \frac{\alpha_x x^2 +\alpha_y y^2 +\alpha_z z^2}{2z_0^2} V_{AC}^0\cos(\Omega t),
\end{equation}
where $x,y,z$ are distances from the minimum along trap axes $X,Y,Z$~(See Fig.~\ref{experiment}.a), and $\alpha_{x,y,z}$ and $z_0$ are parameters dependent on the trap geometry. For our system $z_0=2.19$ mm as follows from electrostatic simulations, which are described in full detail in Ref. ~\cite{JoycePreparation}.
\\
\indent
It can be seen from symmetry considerations (See Fig.~\ref{experiment}.a) that motion in the $X$ and $Y$ directions is non-degenerate only if there is some intentional trap asymmetry introduced to the system. This asymmetry is crucial for our experiments, since we rely on the ability to resolve all three degrees of translational motion present in the optical signal coming from a single detector. The splitting of the degeneracy is done by making a slot in the outer electrode, which allows us to distinguish motion along each of the trap axes by its frequency.
\\
\indent
In the pseudopotential approximation, the particle motion has three main frequencies of oscillation~\cite{Paul1990}:
\begin{equation}
\label{omegaeigen}
\omega_{x,y,z}= |\alpha_{x,y,z}|\frac{q}{m}\frac{V_{AC}^0}{ \sqrt{2}\,z_0^2\,\Omega},
\end{equation}
where $q$ and $m$ are the charge and mass of the particle. The pseudopotential approach is applicable when $\omega_x,\,\omega_y,\,\omega_z<<\Omega$, conditions which are true for our experimental parameters of $\omega_{x,y,z}/2\pi$ and $\Omega/2\pi$ which lie in the range of 150-1000 Hz and 15-30 kHz respectively.
\\
\indent
Charge to mass ratio values of about 10-100 $C \cdot kg^{-1}$ are typical for particles trapped using the electrospray technique, and the data presented in the paper are taken on one of these particles. Special care is taken to make sure that only one particle is transferred and studied in the high vacuum chamber. When multiple particles are trapped, the parameters of the trap (voltage and frequency) are varied towards the unstable trapping conditions, resulting in the consecutive expulsion of the particles from the trap. The process is repeated until a single particle is left over, the case that is easily distinguishable visually via the camera. To minimize charge loss of the particle, laser power for all experiments was limited to 1 mW or lower. Discharge observed at higher laser powers may be related to evaporation of the solution residue left on particles after their preparation~\cite{Hernandez2008}. A 671 nm circularly polarized laser propagating along the $Z$ axis of the trap was used to stabilize the rotation of the particle, and minimize brightness fluctuations attributable to nanoflake reorientation~\cite{Kane2010}. In the presence of the circularly polarized laser, the flake lies in the $X-Y$ plane and is spinning around the axis parallel to the $Z$ axis of the trap.
\\
\indent
Motion of the trapped particle was tracked using a 532 nm linearly polarized laser beam~($\vec{E}\mathbin{\|}Y$) with a width of 0.16 mm~(See Fig.~\ref{experiment}.b). The light scattered from the particle is focused by a lens with a 40 mm diameter located at a distance of 12.5 cm from the center of the chamber and is afterwards divided by a 50:50 beam splitter. One of the split beams is used for rough particle positioning with a CCD (charge-coupled device) camera and provides information about average position of the particle and amount of scattered light at a frame rate of 1 sec$^{-1}$. The second beam falls on a knifeedge of a 90$^\circ$ prism mirror beam splitter at the lens' focal point. This directs light to one of two avalanche photodiodes ($A$ and $B$ in Fig.~\ref{experiment}.c), depending on the position of the particle. The difference between the photodetector signals, $S=\frac{A-B}{A+B}$, can be used to study the center of mass motion of the trapped particle. Here we intentionally choose to use the normalized difference of signals in order to minimize the signal dependence on the laser power and on variations of intensity of scattered light.
\\
\indent
Only components of motion that have non-zero projection onto the image plane in a direction perpendicular to the prism knifeedge will contribute to the signal $S$. Indeed, orienting the prism edge orthogonally to the $Z$ axis of the trap~(see Fig.~\ref{experiment}.c, where prism is shown as viewed along the deflected light direction from the viewpoint at the center of the trap) will maximize the amount of motion detected from the particle oscillations in the $Z$ direction. Similarly, rotating it by 90 degrees to the direction parallel to the $Z$ axis will zero the signal from the oscillations along $Z$ and maximize the $X$ and $Y$ projections, albeit at amplitude $1/\sqrt{2}$ lower than amplitude along $Z$ in the previous case. A rotation angle of $\sim\,$54.7 degrees is chosen in order to make scaling factors for contributions from each of the frequencies to be the same. In this configuration, the fast Fourier transform (FFT) of the signal $S$ contains all three peaks, one for each of the eigenmodes of motion~(Fig.\ref{PowerSpectrum}).
\\
\indent
The signal $S$ is sent to an FPGA (Field-Programmable Gate Array) for digital filtering and processing as well as for generation of slow feedback (cut-off frequency of 4.4 Hz) that is used to keep $A=B$ by adjustment of lens' position on a piezostage. The speed of this feedback is set to be much slower than the typical eigenfrequencies of the particle motion to avoid any cross-talk between the feedback and the motion.
\\
\indent
The second and main purpose of the FPGA is to generate a feedback signal that is fed to the signal generator as an amplitude modulation (AM) input. The feedback signal $S_{feed}= g S \dot{S}$ is created by multiplying the filtered signal $S$ by its derivative~\cite{Novotny2012}, where $g$ is the gain setting of the feedback. For sinusoidal signals, this operation creates an output at \textit{twice} the particle oscillation frequency. The feedback modulation is applied to the trap potential $V_{AC}=V_{AC}^0(1+G_{AM} S_{feed})$, where $G_{AM}$ is the AM value set at the signal generator.
\\
\indent
Once the $p$ is lowered below 1 mTorr, the cooldown of the particle motion is observed if proper feedback conditions are chosen. We have used AM setting of $30\%$ at feedback gain of $g=2.4\times 10^{-4}$ s to show the cooling dependence on $p$. At these parameters, all three degrees of particle motion~(Fig.\ref{PressureAMVoltage}.a) reach $T\sim 20$ K at $p = 4\times 10^{-7}$ Torr. In the region between $10^{-3}$ and $10^{-7}$ Torr, $T$ is strongly affected by the pressure, and even changing $p$ from $10^{-6}$ Torr down to $4\times 10^{-7}$ Torr improves cooling from $T\sim 40$ K down to $T=20$ K.
\\
\indent
To prove that the AM setting of $30\%$ chosen for the experiment is optimal and to show that parametric feedback leads to cooling, we measure the $T$ dependence on the AM setting. It can be seen in Fig.\ref{PressureAMVoltage}.b that higher AM values result in better cooling and there is a transient region, where cooling efficiency is different for each of the degrees of freedom. It is not currently clear why this is the case and more experiments are required before making any statements about the source of this behavior. At the same time, one can notice that increasing the AM setting to values higher than $30\%$ does not improve the cooling, which makes $G_{AM} = 30\%$ the best value for experiments. Absence of further cooling for larger AM values indicates that noise present in the feedback is likely limiting cooling.
\\
\indent
The sensitivit*y of cooling to the feedback noise is enhanced by any non-zero DC electric fields that can be present in the chamber~(arising, for example, from patch potentials~\cite{Harter2014,Wineland1998}). This makes it necessary to balance out their contribution by nulling the total DC field acting on the particle. The cooling optimization is done by finding the settings on auxiliary electrodes (see Fig.~\ref{experiment}.b) that correspond to minimum $T$ of the particle. Although these settings tend to change with time~\cite{Harter2014}, the scale of their drift is on the order of days. Therefore, their values need to be found only once during the measurements. Typical dependence of cooling on electrodes' voltage in the vicinity of the nulling settings is presented in Fig.\ref{PressureAMVoltage}.c. For simplicity of discussion, the figure shows what happens when only voltage $V_3$ is offset from its nulling value, but similar voltage dependencies are observed for the two other electrodes. The presence of a distinct minimum in the $T$ data proves that we have indeed found the voltage settings that minimize the effect of DC fields on the particle motion. The cooling $T$s at the minimum are $T_x=14\pm 8$~K, $T_y=7\pm 4$~K and $T_z=16\pm 5$~K for $X$, $Y$ and $Z$ degrees of motion correspondingly.
\\
\indent
Additional characterization of the flake, including direct measurement of $T$, can be done if we can extract its mass and charge values. The estimation of these parameters is complicated since they usually enter physical equations as a ratio (e.g.~see Eq.~(\ref{omegaeigen})). The coupling of charge and mass can be overcome if some process that affects only charge or only mass exists (or is created) in the system. One of the measurements allowing this is the discharge of the particle that is observed if large enough laser power is applied.
\begin{figure}
 \includegraphics[width=\linewidth]{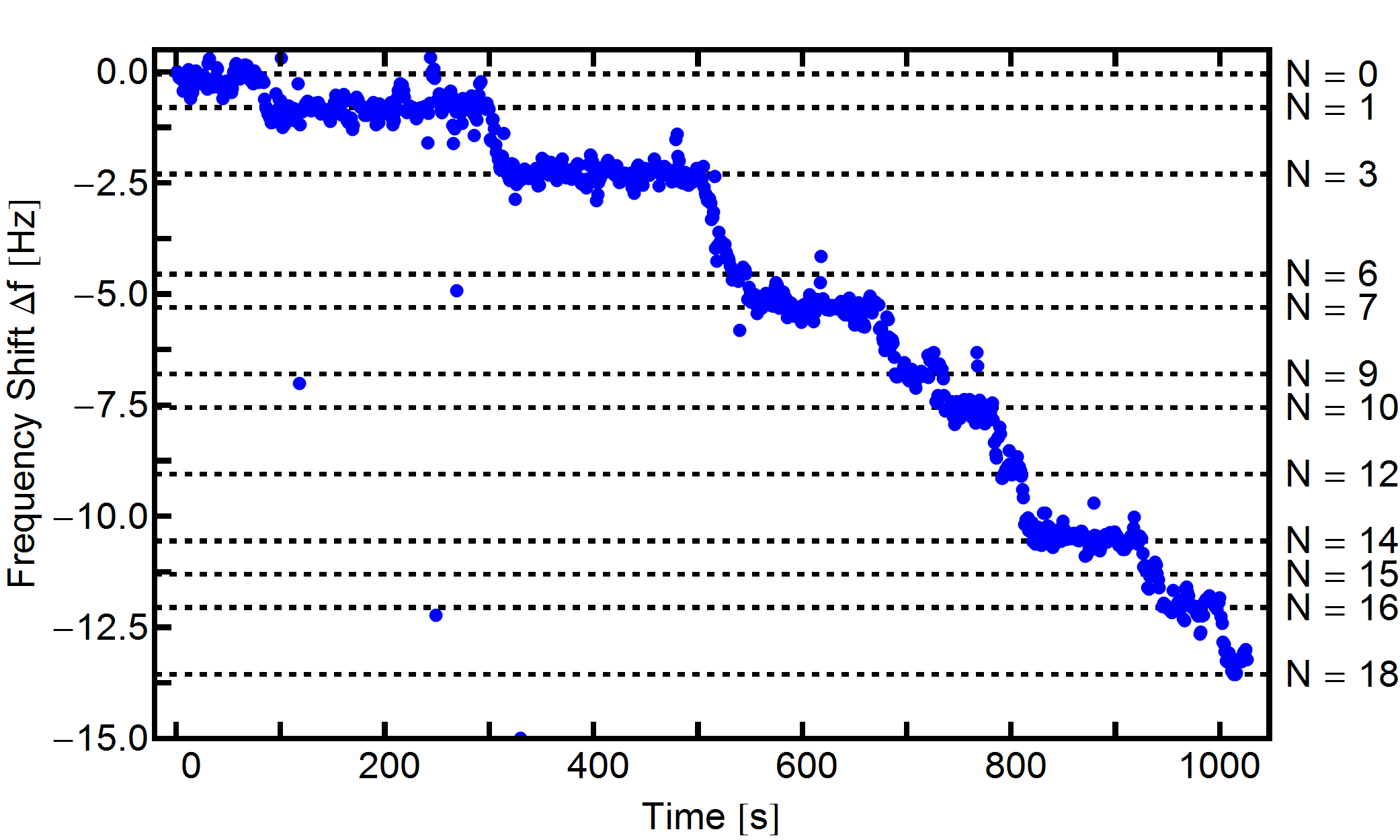}
\caption{\label{Discharge} Time dependence of the particle motion frequency for $Z$ direction. Quantized charge jumps found by fitting the jump step to give integer jump values are shown by dashed lines [$N$ corresponding to each jump is shown on the right].}
\end{figure}
\\
\indent
The discharge sequence at laser power of 1.2 mW with a distinct discharge quantization is shown on Fig.~\ref{Discharge}. While only dependence of frequency of motion along the $Z$ axis on time is presented, jumps in frequency consistent with the depicted ones are observed for motion along $X$ and $Y$ axes. We can find $\delta\omega$ that corresponds to the minimal possible jump (when charge of the particle changes from $N\,e$ to $(N \pm 1)\,e$) by fitting each of the frequency jumps to be equal to $N\delta\omega$, where $N$ is integer. As follows from Eq.~(\ref{omegaeigen}), mass of the particle can be expressed via $\delta\omega$:
\begin{equation}
m = \frac{\delta q}{\delta\omega}\frac{V^0_{AC}}{\sqrt{2}\,z_0^2\,\Omega},
\end{equation}
where $\delta q = e$\cite{Gerlich2001,Gratta2014,Bell2014}. This expression gives us an estimate of $m\approx 1.4\times 10^{-17}$ kg for the particle we present data on. Total charge $q\approx +970\,e$ is found from the ratio $q/\delta q = \omega_z/\delta\omega_z$, where $\omega_z/2\pi = 730$ Hz at the beginning of discharge. It should be pointed that the estimates are given assuming that only charge is changing during the frequency jumps. This assumption is based on the frequency behavior expected from Eq.~(\ref{omegaeigen}). In high vacuum we expect mass to get only smaller since possible changes are limited to desorption of volatiles stuck to the particle surface and to falling off of loosely attached layers. The resulting increase of the eigenfrequency would be contrary to the observed time dependence.
\\
\indent
Further improvement in the cooling and control of the particles can be gained by using phase-locked loops (PLL) for separate cooling of each degree of motion. This narrows the bandwidth of the noise that can couple to the feedback during the feedback generation. While the bandwidth of the current experiment is limited by the necessity to keep the signal from all eigenfrequencies, for PLL the bandwidth of the loop can be made as small as the width of an individual eigenfrequency peak~\cite{Roland2003}. In high vacuum conditions this leads to bandwidth smaller than 1 Hz and a drastic improvement of the signal to noise ratio is predicted if PLL feedback is implemented. PLL will also eliminate noise coupling related to the presence of nonzero electric fields. However, it should be pointed out that the PLL feedback will have its own challenges, since every time a discharge event occurs, the frequency of the oscillations may move out of the PLL bandwidth range, and during the required reset of PLL a temporary lack of proper cooling can lead to the loss of the particle.
\\
\indent
The ability to extract the mass of the particle through its discharge allows for a separate estimate of particle's $T$ if a proper calibration of signal $S$ in terms of spatial displacement is done. However, it should be pointed out that there is an intrinsic nonlinearity in our signal detection that makes the $T$ extraction potentially inaccurate and proper calibration is dependent on exact value of the reflectivity of the particle as well as high signal to noise ratio.
\\
\indent
In conclusion, we have presented experimental results on cooling and stabilization of the graphene nanoplatelets in high vacuum conditions. It was shown that at $p$ of $4\times 10^{-7}$ Torr the particle can stay trapped on a time scale longer than a few days and cooled down below 20 K in all three degrees of motion, reaching particle localization below $1\,\mu m$ in the cooled state. Achieving stable motion of the particle allows us to observe the discharge of the particle and to extract its mass and charge. Finally, it was shown that the elimination of stray fields was necessary to achieve optimal cooling.

This work was supported by the Laboratory for Physical Sciences.
\bibliographystyle{ieeetr}

\end{document}